\def\cm{cm$^{-1}$}
\def\efa{Eu\-Fe$_2$\-As$_{2}$}
\def\bfa{Ba\-Fe$_2$\-As$_{2}$}
\def\BaFeCoxAs{Ba\-(Fe$_{1-x}$\-Co$_x$)$_2$\-As$_{2}$}

\def\bfca{Ba\-(Fe$_{0.92}$\-Co$_{0.08})_2$\-As$_{2}$}
\def\bfna{Ba\-(Fe$_{0.95}$\-Ni$_{0.05})_2$\-As$_{2}$}
\def\bfma{Ba\-(Fe$_{1-x}$\-$M_{x})_2$\-As$_{2}$}
\def\bkfa{Ba$_{0.6}$K$_{0.4}$Fe$_2$As$_2$}

\def\tc{$T_{c}$}

\documentclass[prb,twocolumn,showpacs,amsmath,amssymb]{revtex4}
\usepackage[dvips]{graphicx}
\usepackage{graphicx}

\begin{document}

\title{Electrodynamics of electron doped iron-pnictide superconductors: \\
Normal state properties}

\author{N. Bari\v{s}i\'{c}}
\author{D. Wu}
\author{M. Dressel}
\affiliation{1.~Physikalisches Institut, Universit\"at Stuttgart, Pfaffenwaldring 57, 70550 Stuttgart, Germany}
\author{L. J. Li}
\author{G. H. Cao}
\author{Z. A. Xu}
\affiliation{Department of Physics, Zhejiang University, Hangzhou 310027, People's Republic of China}

\date{\today}

\begin{abstract}
The electrodynamic properties of \bfca\ and \bfna\ single crystals have been investigated by reflectivity measurements in a wide frequency range. In the metallic state, the optical conductivity consists of a broad incoherent background and a narrow Drude-like component which determines the transport properties; only the latter contribution strongly depends on the composition and temperature. This subsystem reveals a $T^2$ behavior in the dc resistivity and scattering rate disclosing a hidden Fermi-liquid behavior in the
122 iron-pnictide family. An extended Drude analysis yields the frequency dependence of the effective mass (with $m^*/m_b\approx 5$ in the static limit) and scattering rate that does not disclose a simple power law. The spectral weight shifts to lower energies upon cooling; a significant fraction is not recovered within the infrared range of frequencies.
\end{abstract}

\maketitle

\section{Introduction}

The family of the recently discovered iron-pnictides surprises not only because of its extreme electronic properties, but also because of its fine-tuned interplay of electronic, spin and lattice degrees of freedom. The investigation is facilitated by the possibility to vary the systems by chemical substitution in many ways; often changing simultaneously structure and carrier density. However, the important parameters and their influence are not understood by now: in fact, it is not even obvious that the core of the problem has been identified.\cite{Xu08,Tesanovic09,Ishida09} Such or similar problems are common to unconventional superconductors. In the case of cuprates, it is debated for long how a Mott insulator becomes conducting upon charge-carrier doping, how important the coupling to neighboring planes is, how magnetic and superconducting orders develop, what the influence of fluctuations is on the superconducting ground state and many other issues which remain unresolved. In the case of heavy-fermion materials, the hybridization of localized $f$ electrons and conduction electrons causes unusual properties that are subject to intense research even for longer than cuprates.\cite{Ott87} These intermetallic compounds are usually good metals with strong influence of local electron-electron interaction leading to magnetic order at low temperatures; the tuning is not done by doping but by changing the bandwidth compared to the interaction strength. Organic superconductors are even another class of correlated electron matter, which are distinctively different due to their layered structure of molecules with $\pi$ electrons; nevertheless with similar issues relevant  for the understanding of superconductivity evolving out of a correlated electronic state.\cite{Fukuyama06} While a common understanding of the overall behavior in those materials has surfaced over the years, the subtleties remain a challenge, and novel surprising facts are discovered regularly.

In the case of iron-pnictides, the parent compounds are semimetals where the conduction electrons order in an antiferromagnetic fashion forming a collinear spin-density-wave (SDW).\cite{Cruz08,HuangQ08} A structural phase transition precedes the SDW transition at a slightly higher temperature.\cite{ Rotter08} The SDW instability is usually associated with the imperfect interband nesting of itinerant electrons. Doping by electrons or holes suppresses both the structural transition and the magnetic order rapidly and superconductivity is found with \tc\ above 50~K.\cite{Chu09} However, since the parent iron pnictide compounds are already metallic, the increase in the charge carrier concentration as a result of doping does not seems to be a decisive factor in the development of superconductivity. Physical or chemical pressure causes superconductivity as well, indicating that the influence of the structural arrangement cannot be overestimated. The enormous and inverse isotope effect makes the situation even more puzzling.\cite{Liu09,Shirage09,Yanagisawa09}

Band structure calculations agree with angle resolved photoemission spectroscopy (ARPES) indicating that several electronic bands are close to the Fermi level, not all of them well separated in energy and momentum.\cite{Zabolotnyy2009,Yi2009,Terashima09} This considerably complicates the understanding of these materials. At this point it is not clear how to associate the observed electron correlations with the multiband structure of iron-pnictides.\cite{Qazilbash09}
Most experimental methods, in accord with theory,\cite{Mazin08} evidence a fully gapped superconductor with no nodes of the order parameter. Its symmetry might be $s_{\pm}$ wave, i.e.\ reverses sign for electron and hole pockets of the Fermi surface. ARPES results\cite{Evtushinsky09, Terashima09} yield two gaps --~different in energy by a factor of 2~-- in certain parts of the Fermi surface which evolve simultaneously below $T_c$.

Optical investigation turned out to be one of the most efficient and powerful tools to reveal the electronic and dynamic properties of multiband conductors, iron-pnictides in particular.\cite{Chen08,Chen09,Dong08,Hu08,Akrap09,Nakajima09,Li08SC,Hu09c,Wu09,Wu09NP,BFilm10,Bernhard09,Dirk09,Homes10} Here we will restrict ourselves to the 122 family for which single crystals of suitable size are readily available. Direct observation of the partial SDW gap was first achieved in SrFe$_2$As$_2$ and \bfa\ and interpreted by a two-gap structure.\cite{Hu08, Akrap09,Nakajima09} This, however, seems to be controversial since in \efa\ \cite{Wu09} and CaFe$_2$As$_2$ \cite{Nakajima09} and in a more recent study on \bfa\ \cite{Pfuner09} similar features have not been reported.\cite{remark1} Chen {\it et al.} \cite{Chen09} performed optical experiment on BaNi$_2$As$_2$ which undergoes a structural phase transition at 130~K \cite{Ronning08,Sefat09} indicates the opening of a gap. However, so far no evidence for a magnetic phase was reported. Interestingly, in contrast to the SDW transition in BaFe$_2$As$_2$ or \efa, the spectral weight of BaNi$_2$As$_2$ shifts to much higher frequencies (well above 10\,000~\cm). The extended Drude analysis of the complex
conductivity of \efa\ yields a linear behavior of the
frequency-dependent scattering rate below $T_{\rm SDW}$,
indicating an interaction between the charge carriers and spin
fluctuations in the spin-density-wave state.\cite{Wu09} From
investigating the normal state properties of the superconducting
Ba$_{0.55}$K$_{0.45}$Fe$_2$As$_2$ and by applying a sophisticated
model, Yang {\it et al.}\cite{Yang09} also concluded that the
electron system couples to bosonic excitations. In their analysis it
became obvious that it is important to separate the different
contributions to the optical conductivity.

In Ref.~\onlinecite{Wu09NP} we briefly reported on our dc resistivity and optical measurements in a broader context of the
observed universal electronic behavior in \bfma\ single crystals.
Here we will give a full presentation of our optical experiments
and thorough discussion of our results in the normal state. A
proceeding paper (Ref.~\onlinecite{PartII}) will deal with the
superconducting state.

\section{Experimental Details}

Single crystals of \bfca\ and \bfna, i.e.\ close to optimal
doping,  were grown using FeAs as self-flux dopants
\cite{Shuai,Chu09,Li09} and characterized by x-ray,
EDX-microanalysis, transport and susceptibility measurements. The
platelets with a typical size of $4~{\rm mm}\times 2~{\rm
mm}\times 0.5~{\rm mm}$ have naturally flat surfaces. The
temperature-dependent dc resistivity was obtained by standard
four-probe technique, using silver paint as contacts. Samples were characterized by measuring the magnetic susceptibility down to 4~K by a Quantum Design SQUID system. Applied magnetic field of 5 Oe was perpendicular to the $ab$ plane.\cite{Wu09NP}

The temperature dependent optical reflectivity $R(\omega,T)$ (in $ab$ plane) was measured in a wide frequency range from 20 to 37\,000~\cm\ using a coherent-source spectrometer in the THz range ($20-40$~\cm),\cite{Gorshunov07,Dressel08} two infrared Fourier transform spectrometers (Bruker IFS 66v/s and IFS 113v) ($30-12\,000$~\cm) and a Woollam variable-angle spectroscopic ellipsometer extending up to the ultraviolet  ($6000 - 37\,000$~\cm, restricted to room temperature). Part of the experiments were performed by employing an infrared microscope Bruker Hyperion. Using different CryoVac and homemade He cryostats we could continuously cool down to approximately 10~K. As reference we used either freshly evaporated aluminum or gold mirrors; alternatively gold was deposited onto the sample {\it in situ} and the measurements repeated at each temperature. While the accuracy in the absolute value might not be better than 1\%, the uncertainty in the relative change with temperature is significantly better. The low-frequency extra\-polation was done according to the dc conductivity measured on the same specimens.
In the superconducting state we have smoothly extrapolated
$R(\omega)$ without introducing artificial kinks or steps.
The complex optical conductivity $\hat{\sigma}(\omega)=\sigma_1(\omega)+{\rm
i}\sigma_2(\omega)$ was calculated from the reflectivity spectra
using Kramers-Kronig analysis.\cite{DresselGruner02}

\section{Electrical resistivity}
\label{sec:resistivity}

\begin{figure}
 \centering
 \includegraphics[width=0.7\columnwidth]{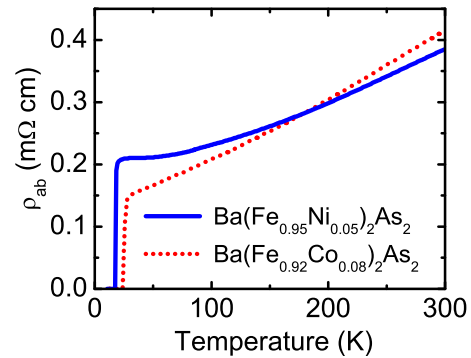}
 \caption{\label{fig:Fig1} (Color online) Temperature dependence of the electrical resistivity $\rho_{ab}(T)$ measured in the $ab$-plane of \bfca\ and \bfna.}
\end{figure}

The dc resistivity is measured to characterize our pnictides
single crystals and to determine the energy scales which enter the
problem. On one hand, resistivity is a bulk probe and therefore
quite sensitive to extrinsic disorder or inclusion phases. On the
other hand, it is a property that results from integration over
the whole Fermi surface, and it is therefore sensitive to most of
changes in the intrinsic behavior of the electronic system.
Several samples of each compound are measured, showing temperature
dependences identical to those presented in Fig.~\ref{fig:Fig1}.
The high data reproducibility indicates that samples are single
phased and together with the observation of a  sharp
superconducting transition of $\delta T_c\approx 1$~K suggest that
they are uniformly doped and of good quality.\cite{Barisic08}

The absolute value of the room-temperature dc resistivity is $\rho_{ab}=415~\mu\Omega{\rm cm}$ and $385~\mu\Omega{\rm cm}$ for \bfca\ and \bfna, respectively, indicating a fairly good metal.\cite{remark4} Indeed, by lowering the temperature from 300~K down to the superconducting transition at $T_c=25$ and 20~K, the resistivity decreases monotonously. Following previous attempts, we could fit equally the temperature dependence of the resistivity by the power-law
\begin{eqnarray}
\rho(T)&=&\rho_0 + AT^n \quad ,
\label{eq:Tn}
\end{eqnarray}
with $n=1.26$ for \bfca\ and $n=1.55$ for the Ni compound, or by the sum of a linear term and a quadratic term
\begin{eqnarray}
\rho(T)&=&\rho_0 + A_1T + A_2T^2 \quad ,
\end{eqnarray}
with $A_1=7 \times 10^{-7}~\Omega {\rm cm K}^{-1}$, $A_2=8.6 \times 10^{-9}~\Omega {\rm cm K}^{-2}$ for \bfca\ and $A_1=4 \times 10^{-7}~\Omega {\rm cm K}^{-1}$, $A_2=1.5 \times 10^{-9}~\Omega {\rm cm K}^{-2}$ and \bfna.

A quadratic temperature dependence is expected for electron-electron scattering that is supposed to be dominant for correlated electron systems at low temperatures and subject to Landau's theory of a Fermi liquid.\cite{Abrikosov63,Pines66}
The observation of a non-Fermi liquid exponent ($n\neq 2$) or linear temperature dependence of the resistivity is often taken as the hallmark of a quantum critical regime.\cite{Sachdev01,Lohneysen07} Similar behaviors were observed and recently discussed in heavy-fermions systems, organic superconductors or in high-temperature superconductors.\cite{Doiron09} Quantum critical phenomena were initially investigated using a scaling analysis, based on an extension of the Landau theory of phase transitions, \cite{Landau} which -- in the original version -- describes the classical critical behavior near a second-order phase transition at $T_c$. Theories for the continuous phase transitions give the so-called scaling laws.\cite{Fisher67,Kadanoff67}  The corresponding low-temperature dependences of the resistivity, for the spin-fluctuation models, result --~depending on the dimensionality of the system~-- in the power-law coefficient which between 1 and 1.5 in the case of antiferromagnetic fluctuations.\cite{Hertz76,Millis93,Moriya95}

\begin{figure*}
 \centering
\includegraphics[width=8cm]{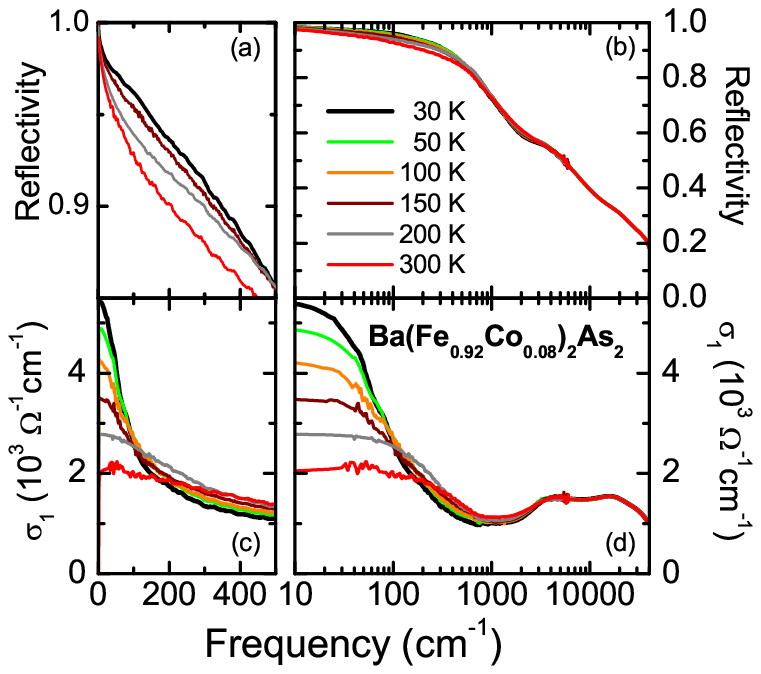}
\hfill
\includegraphics[width=8cm]{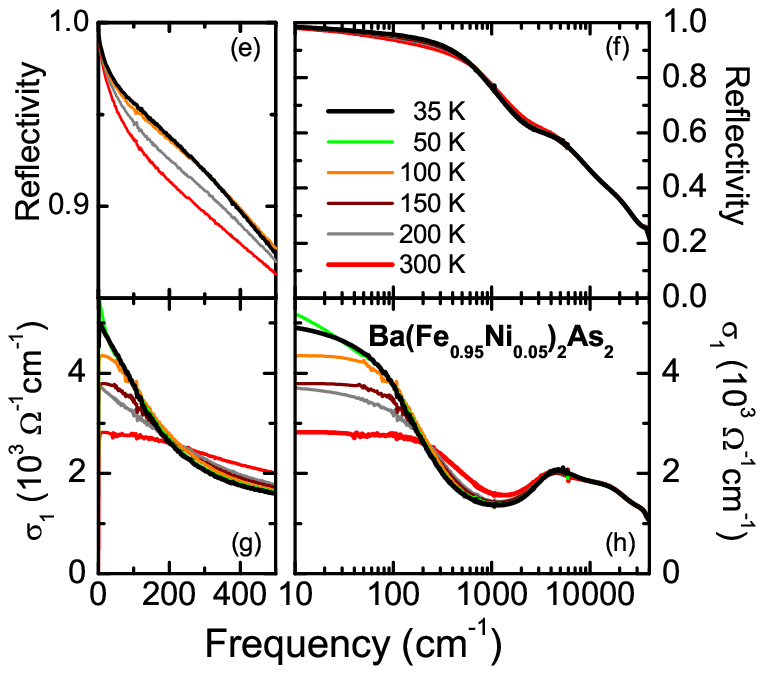}
 \caption{\label{fig:Fig2} (Color online)
Optical properties of \bfca\ (left frame) and \bfna\ (right frame) measured at different temperatures $T>T_c$ as indicated. The upper panels show the normal state reflectivity. Panels (a) and (c), (e) and (g) are linear-scale enlargements of the low-frequency reflectivity and conductivity at $T=300$, 200, 100 and 30 (35)~K (from the bottom to the top). The optical conductivity plotted in the lower panels is obtained from the Kramers-Kronig analysis of the corresponding reflectivity.}
\end{figure*}

Thus it is not surprising that the proximity of the magnetic phase and the observation of the non-Fermi Liquid behavior initiated arguments that the superconducting ground state is a consequence of an underlying quantum criticality.\cite{Chu09,Gooch09,Jiang09}

Based on our analysis of the optical spectra of a large number of iron-pnictides,\cite{Wu09NP} we suggest an alternative approach. Instead of analyzing the data using Matthiessen's rule, where different scattering contributions are treated additively, we introduce subsystems which add up their contributions to the conductivity. As will be discussed below in Sec.~\ref{sec:Hidden}, our findings suggest a two-component metal with a temperature-independent background ($\sigma_B$) and an essentially Fermi-liquid channel ($\sigma_{N}$) that determines the electronic properties of the pnictides.

\section{Reflectivity and optical conductivity}
\subsection{Overall behavior}

The electrodynamic properties of both compounds under investigation are very similar, which can serve as an indication that the behavior is intrinsic for the normal state of all members of the 122-family. The optical reflectivity $R(\omega)$ exhibits a metallic behavior at all temperatures $T>T_c$. It drops almost linearly with frequency up to more than 1000~\cm\ where it changes slope around $R=0.6$ and continues to decrease again linearly up to the visible. In the logarithmic plots of Figs.~\ref{fig:Fig2}(b) and (f), this becomes a shoulder-like feature which has been previously reported in the 122 family.\cite{Wu09} It is attributed to an interband transition.\cite{Singh08,remark2} Significant changes with temperature are seen in the far-infrared, where $R(\omega)$ rises by decreasing temperature as depicted in Figs.~\ref{fig:Fig2}(a) and (e). Correspondingly, the mid-infrared reflectivity exhibits the opposite temperature behavior: $R$ rises with the temperature; the isobetic point in $R(\omega)$ is at $\omega_{i}/(2\pi c)\approx 1150$~\cm\ and 700~\cm, for \bfca\ and \bfna, respectively.

In Figs.~\ref{fig:Fig2}(d) and (h) the corresponding optical conductivity of \bfca\ and \bfna\ is plotted for different temperatures in the metallic state. As $T$ drops, a Drude-like contribution evolves in accord with the rising dc conductivity. Its width decreases on reducing temperature; a few degrees above $T_c$, at $T=30$ and 35~K, the scattering rate is around $100-200$~\cm\ (for a more detailed analysis, see below). In the range of $500 - 2000$~\cm\ the conductivity decreases upon cooling, implying that the spectral weight is significantly reduced in this frequency range as the temperature is decreased. As discussed in the subsequent Section, part of it shifts to lower frequencies; however a sizeable fraction is not recovered within the spectral range considered here. For both compounds the conductivity around 1000~\cm\ reaches a more-or-less temperature independent local minimum with a value of $\sigma_1\approx 1000~(\Omega {\rm cm})^{-1}$.

As mentioned above, the shoulder in the reflectivity plot corresponds to an interband transition that peaks around 4400~\cm\ and a second near-infrared feature around  15\,800~\cm. The mid-infrared maximum is more pronounced for \bfna\ compared to the cobalt analogue.

\subsection{Spectral weight analysis}
\label{sec:SpectralWeight}
\begin{figure}
 \centering
\includegraphics[width=\columnwidth]{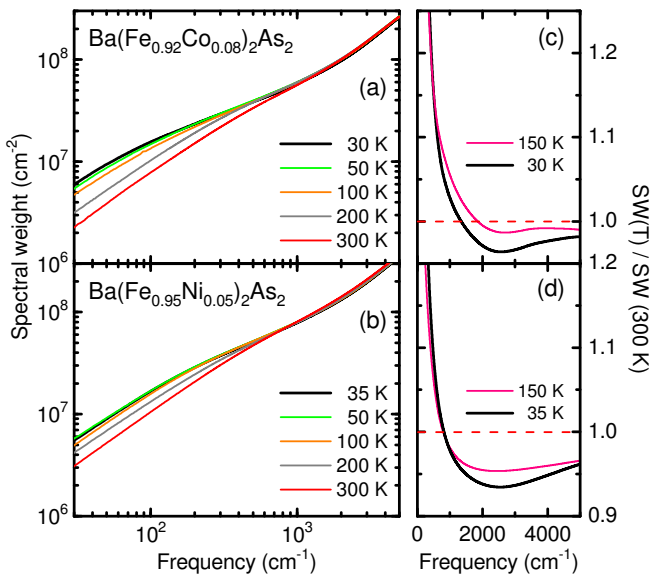}
 \caption{\label{fig:Fig3} (Color online)
Frequency dependence of the integrated spectral weight for (a) \bfca\ and (b) \bfna\ at different temperatures $T>T_c$. The development of the spectral weight SW$(T)$ normalized to their room temperature value SW(300~K) is plotted in panels (c) and (d) for the two compounds.}
\end{figure}

In order to get a more quantitative picture of the spectral weight redistribution with temperature, in Fig.~\ref{fig:Fig3} we plot the integrated conductivity \cite{DresselGruner02}
\begin{equation}
{\rm SW}(\omega_c) = 8\int_0^{\omega_c}\sigma_1(\omega)\,{\rm d}\omega = {\omega_p^2}=\frac{4\pi~Ne^2}{m}
\label{eq:spectralweight}
\end{equation}
as a function of cut-off frequency $\omega_c$, where $N$ is the
density of carriers participating in optical transitions below the
frequency $\omega_c$. Up to $10^4$~\cm\ no obvious step or
saturation evidences a plasma frequency; this is in accord with
the gradual decrease in the reflectivity $R(\omega)$ to higher
frequencies [Fig.~\ref{fig:Fig2}(b)]. With temperature dropping
from $T=300$ to 35~K, the spectral
weight around 2500~\cm\ is reduced by approximately 5\%\ for both
compounds, an effect which is significant and beyond the
uncertainty of our experiments. To elucidate this point, we have
normalized the spectral weight to its room temperature behavior:
SW($\omega,T$)/SW($\omega,300~{\rm K}$) and plotted the ratio in
Figs.~\ref{fig:Fig3} (c) and (d) for an intermediate ($T=150$~K)
and low temperature (30~K, 35~K). The overall behavior is very similar for both compounds. As $T$ is reduced, there is an accumulation of spectral weight at low frequencies that is partially compensated in the range of $500 - 2000$~\cm, leading to SW($\omega,T$)/SW($\omega,300~{\rm K})<1$. Albeit the ratio goes through a minimum around 2500~\cm\ and increases for higher frequencies, the relative spectral weight does not reach unity within the infrared range. As the temperature is lowered, the spectral-weight loss gets more pronounced. The surprising fact that the spectral weight is not recovered up to 5000~\cm\ is a strong argument that correlation effects are of
importance, as suggested previously.\cite{Qazilbash09,Chen09c} At this
point, we cannot make well-founded statements on the recovery of
the spectral weight in the energy range of the interband
transitions, since our ellipsometric measurement were confined to
ambient temperature. The necessary experiments are on our agenda.

To determine the plasma frequency, we have chosen $\omega_c/(2\pi c)=2500$~\cm\ as a suitable cut-off frequency. This leaves out the high frequency interband transition which starts to dominate the spectral weight above 2500 \cm\, and considers mainly the itinerant electrons. According to Eg.3, we obtain $\omega_p/(2\pi c) = 12\,000$~\cm\ for \bfca\ and 15\,000~\cm\ for the Ni analogue. The values are comparable to findings of other groups in  compounds of the 122 family.\cite{Hu08,Li08SC,Hu09c,Pfuner09,Wu09,Nakajima09} From the right hand side of the Eq.~(\ref{eq:spectralweight}) we can estimate the carrier density for \bfca\ and \bfna\ to $N=1.6\times 10^{21}~{\rm cm}^{-3}$ and $2.5\times 10^{21}~{\rm cm}^{-3}$, respectively, assuming $m_b=m_0$.
The values are summarized in Tab.~\ref{table1}.

\subsection{Generalized Drude analysis}

\begin{figure}
 \centering
\includegraphics[width=0.7\linewidth]{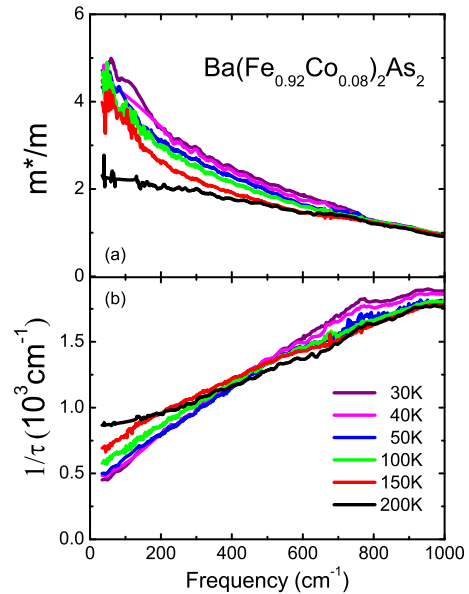}
 \caption{\label{fig:Fig4a} (Color online)
Frequency dependence of (a) the effective mass and (b) the scattering
rate of \bfca\ at different temperatures obtained from the extended Drude analysis of the conductivity data.}
\end{figure}

The extended Drude analysis has matured as a common tool to get insight into the low-energy excitations and the relevant scattering mechanisms of correlated electron systems.\cite{DresselGruner02} The frequency dependent scattering rate $1/\tau(T,\omega)$ and frequency dependent effective mass $m^*(T,\omega)$ is calculated from the complex conductivity spectra which is given by:
\begin{equation}
\hat{\sigma}(\omega)=\frac{\omega_p^2}{4\pi}
\frac{1}{\Gamma_1(\omega)-{\rm i}\omega[m^*(\omega)/m_{b}]} \quad .
\label{eq:gen-drude2}
\end{equation}
Here $\Gamma_1(\omega)$ is the real part of a complex frequency dependent scattering rate $\hat{\Gamma}(\omega)=\Gamma_1(\omega)+{\rm i}\Gamma_2(\omega)$. The imaginary part is related to the frequency dependent mass $m^*/m_{b}=1-\Gamma_2(\omega)/\omega$ enhanced compared to the bandmass $m_b$. From the complex conductivity one can obtain expressions for $\Gamma_1(\omega)$ and $m^*(\omega)$ in terms of $\sigma_1(\omega)$ and $\sigma_2(\omega)$ in following forms:
\begin{subequations}
\label{eq:extendedDrude}
\begin{eqnarray}
\Gamma_1(\omega)&=&\frac{\omega_p^2}{4\pi}
\frac{\sigma_1(\omega)}{|\hat{\sigma}(\omega)|^2}
\label{eq:gam-w}\\
\frac{m^*(\omega)}{m_{b}}&=&\frac{\omega_p^2}{4\pi}
\frac{\sigma_2(\omega)/\omega}{|\hat{\sigma}(\omega)|^2} \quad .
\label{eq:mstar-w}
\end{eqnarray}
\end{subequations}
In Figs.~\ref{fig:Fig4a} and \ref{fig:Fig5} the scattering rate 2$\pi$ $[{\tau(\omega)}]^{-1} = \Gamma_1(\omega)$ and normalized mass $m^*(\omega)/m_b$ are plotted as a function of frequency for various temperatures in the metallic state.

\begin{figure}
 \centering
\includegraphics[width=0.7\linewidth]{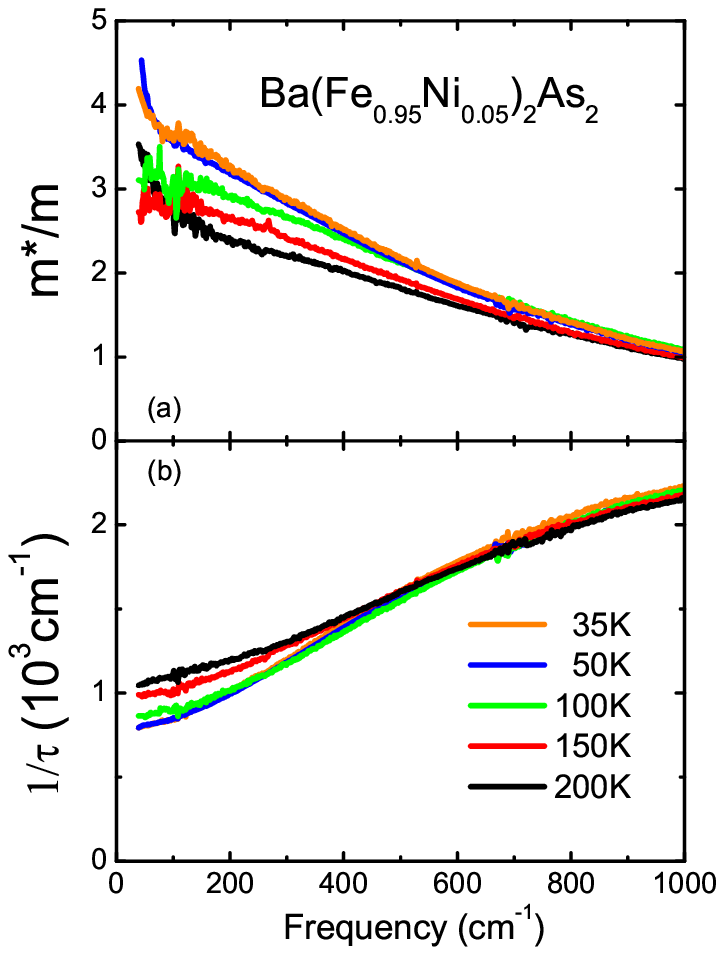}
\vspace*{2mm}
\hspace*{-3mm}
\includegraphics[width=0.69\linewidth]{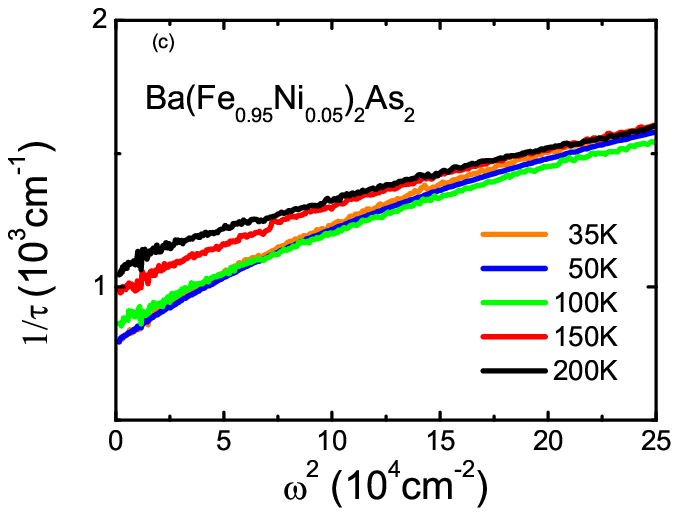}
 \caption{\label{fig:Fig5} (Color online)
Frequency dependence of (a) the effective mass and (b) scattering rate of \bfna\ calculated from the optical spectra for different temperatures.
(c) Scattering rate of \bfna\ plotted as a function of $\omega^2$ for several temperatures.
\label{fig:Fig5} }
\end{figure}

For all measured temperatures, we find (below 800 \cm) a steady increase in the effective mass, indicating deviations from a simple Drude response. The mass enhancement becomes stronger as the temperature is lowered which is usually attributed to correlation effects. For $\omega\rightarrow 0$ it reaches $m^*/m_b=5$ for both materials. It should be noted that the magnitude depends on the reference values of $\omega_p^2$ in Eq.~(\ref{eq:extendedDrude}). As argued above, we have chosen plasma frequencies $\omega_p/(2\pi c)=12\,000$ and 15\,000~\cm.\cite{remark6} Below a certain frequency, which is approximately 750~\cm\ in the case of \bfca\ and 900~\cm\ for \bfna, we find that the effective mass starts to exhibit a temperature dependence.

\begin{table*}
\caption{\label{table1}Transport and optical parameters of \bfca\ and \bfna\
as obtained from our investigations on single crystals.
$\rho_{ab}$ denotes the room-temperature resistivity.
$\omega_p$ is the plasma frequency within the highly conducting
plane, from which the carrier density $N$ and the bandmass $m_b$
is obtained. The effective mass $m^*/m_b$ is derived from the low-frequency limit of the extended Drude analysis. From the room temperature scattering rate $1/\tau$, derived from the dc resistivity,
the mean free path $\ell$ is evaluated, assuming a Fermi velocity
of $v_F= 2\times 10^6$~m/s [typical for metals (Ref.~\onlinecite{Kittel})]. From the temperature dependence of the dc resistivity $\rho_N(T)=\rho_0+ AT^2$ we obtain the prefactor $A$.
}
\begin{ruledtabular}
\begin{tabular}{l cccccccccccc}
{}&$\rho_{ab}$    & $\omega_p/(2\pi c)$ & $N$ & $m_b$
& $m^*$    & $1/\tau$ & $\ell$ & $A$ \\
{}&(m$\Omega$\,cm)& (cm$^{-1}$)         & (cm$^{-3}$) &$m_0$
& $m_b$    & (s$^{-1}$) & (nm) & $\Omega {\rm cm K}^{-2}$\\
\hline
\bfca &  0.415   &  12\,000 & $1.6\times 10^{21}$ & 1 & 5 & $3\times
10^{13}$ & 53 & $6.64\times 10^{-9}$\\
\bfna & 0.385   &  15\,000 & $2.5\times 10^{21}$ & 1 & 4 & $4\times
10^{13}$ & 37 & $5\times 10^{-9}$
\end{tabular}
\end{ruledtabular}
\end{table*}

For \bfca\ the scattering rate reveals a linear frequency dependence at most temperatures [Fig.~\ref{fig:Fig4a}(b)], where the slope increases on decreasing $T$. A similar behavior was reported for the non-super\-conducting parent compound \efa\ and for hole-doped \bfa.\cite{Wu09,Yang09} In general this behavior is assumed to be a signature of bosonic excitations, like spin fluctuation.
However, for \bfna\ the behavior is distinctly different: we can
find a linear dependence of the scattering rate on frequency for
the limited range between 200 and 600~\cm, with the slope
decreasing with $T$ in the same way as it does for \bfca. This,
however, does not hold over an extended range. Also a
$1/\tau\propto \omega^2$ dependence does not describe the entire
frequency dependence satisfactorily, as demonstrated in
Fig.~\ref{fig:Fig5}(c). While for elevated temperatures the
agreement is rather good, the deviations increase as $T$ drops.
Interestingly, a fit by a power law $1/\tau(\omega)\propto
\omega^{n}$ yields an exponent $n=1.4$ over quite some range
similar to what we obtained from the fit of $\rho(T)$ by
Eq.~(\ref{eq:Tn}). This, however, might be a coincidence and
requires additional investigations before entering further discussion.

The reduction of $1/\tau$ when lowering the temperature is
somewhat expected for a metal. More surprising is the fact that
the scattering rate increases above 400~\cm\ as $T$ decreases. A
similar behavior is observed in other strongly correlated electron
systems, like high-$T_c$ superconductors, UPd$_2$Al$_3$, but also
chromium.\cite{Dressel02,Basov02} It was pointed out by Basov {\it
et al.} \cite{Basov02} in the context of high-temperature superconductors that $1/\tau(\omega)$ reproduces the main
features of the density of states including the gap or pseudogap
followed by a  sharp peak. Thus, a shift of weight in scattering
rate from low to high frequencies might be interpreted as a
signature of a gap or pseudogap leading to a reduction in the
electronic density of states. Interestingly, a very similar
behavior is found in the parent compound \efa,\cite{Wu09} where it
was, again, attributed to the development of the SDW gap at low
temperatures. However, from our susceptibility measurements we do
not see any indications of a magnetically ordered state. Also the
dc resistivity does not give any hint of a SDW transition in our electron doped
Ba($M_{x}$Fe$_{1-x}$)$_2$As$_2$.
Similarly, we do not find indications of gap-like features in
the reflectivity data. Furthermore,
the energy scale in frequency and temperature seems to be much to high for a SDW transition. All those arguments indicate that a change in the density of states is not the origin of the unusual increase in scattering rate above 400~\cm. Importantly, it cannot be ruled out, that the shape of $1/\tau(\omega)$ is just a consequence of a more complicated conduction mechanism. Multiple bands may induce a complex form of intra- and inter-band (cross) scattering,  which may even vary with temperature. As a consequence, the single-component analysis --~as performed here~-- leads to features which are of no specific meaning.
Nevertheless, the agreement in $1/\tau(\omega)$ between the parent compounds and the doped iron pnictides for measured frequencies seems unlikely to be a coincidence, since not only the overall behavior but also the energy scales are identical. This, again, can be an indication that, the doping has only small influence on the scattering mechanism.

\begin{figure*}
 \centering
\includegraphics[width=0.7\linewidth]{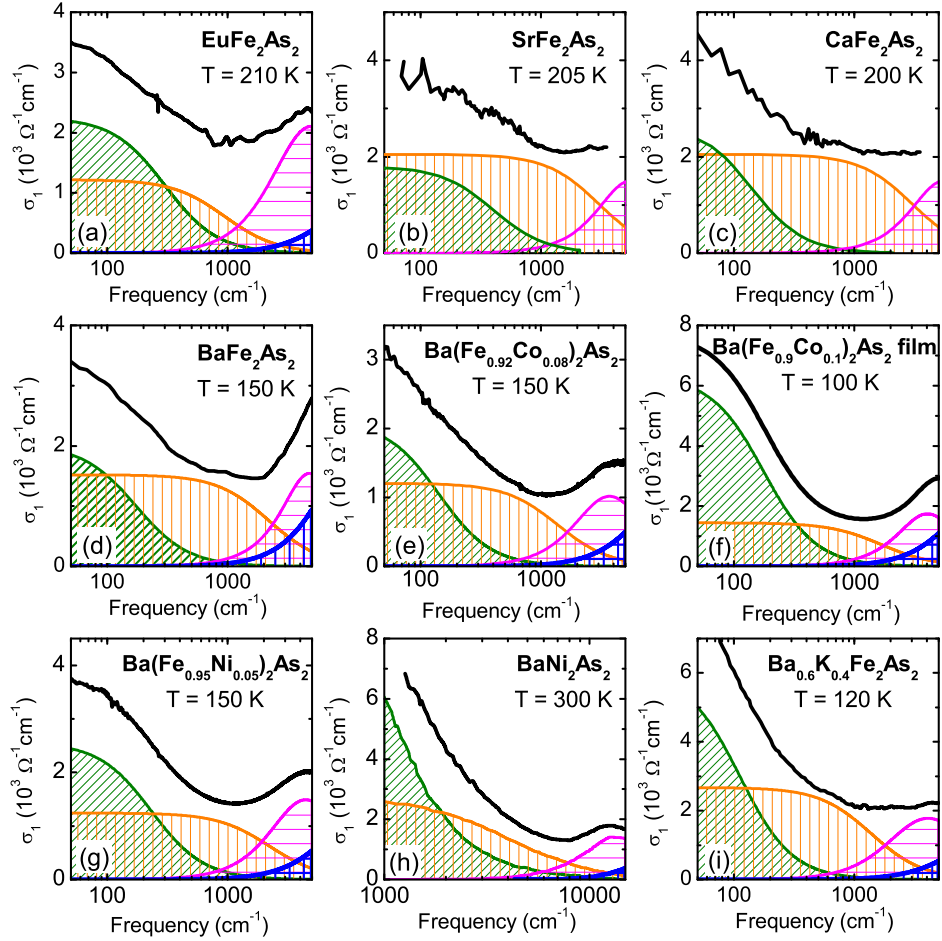}
 \caption{\label{fig:Fig6} (Color online)
In the metallic state the optical conductivity of different iron-pnictides can be described in all known cases by two Drude terms ($\sigma_N$-densely shaded green and $\sigma_B$-medium shaded orange) and two oscillators in the mid-infrared (sparsely hatched magenta and blue). The spectra of the parent compounds at temperatures above the SDW transition: (a)~\efa\ \cite{Wu09}, (b)~SrFe$_2$As$_2$ \cite{Hu08}, (c)~CaFe$_2$As$_2$ \cite{Nakajima09}, and (d) BaFe$_2$As$_2$ \cite{Nakajima09}. For $T>T_c$ the conductivity spectra of the electron doped superconductors (e)~\bfca, (f)~Ba\-(Fe$_{0.9}$\-Co$_{0.1})_2$\-As$_{2}$ film \cite{BFilm10} and (g)~\bfna are very similar. However, the decomposition also holds for the hole doped superconductor (i) Ba$_{0.6}$K$_{0.4}$Fe$_2$As$_2$ \cite{Li08SC}, and (h)~when Fe is completely substituted by Ni\cite{Chen09}, although such materials are more metallic (and thus plotted on a different frequency scale). This latter is reflected in a larger spectral weight of the Drude components.}
\end{figure*}

\section{Discussion}
\subsection{One-component analysis}
From our transport and optical measurement of \bfca\ and \bfna\ single crystals in highly-conducting $ab$-plane, we have deduced several characteristic parameters listed in Table~\ref{table1}. From the spectral weight analysis the carrier density was estimated to $1.6\times 10^{21}~{\rm cm}^{-3}$ and $2.5\times 10^{21}~{\rm cm}^{-3}$ for \bfca\ and \bfna, respectively, assuming a bandmass $m_b\approx m_0$. Taking into account that the volume of the unit cell
$V \approx0.2\times 10^{-21}~{\rm cm}^{3}$,\cite{Rotter08} one find approximately 0.32 and 0.5 itinerant electrons per unit cell, respectively; although there are two Fe ions per unit with a nominal valence of Fe$^{2+}$ meaning six $d$ electrons per Fe atom. A similar value of the charge carrier density was obtained for the parent compound \efa,\cite{Wu09} which --~together with the observation that chemical or physical pressure can also cause superconductivity \cite{Torikachvili08,Jiang09,Ren09,remark7}~-- leads to conclusion that crystal structure rather than carrier doping needs to be fine-tuned to provide the superconductivity. The fact that only a fraction of the Fe electrons contribute to conductivity suggests a distinction of localized and itinerant charge carriers. The effect of local magnetism of the Fe atoms was pointed by Yildirim and others.\cite{Yildirim08,Moon09} In this regard, a comparison with the completely substituted compounds BaNi$_2$As$_2$ might be of interest where the carrier density is much higher.\cite{Chen09} No magnetic order was detected and no superconductivity, but a structural distortion which causes a slight reduction in conduction carriers.

The extended Drude analysis shown in Figs.\ref{fig:Fig4a}(b) and \ref{fig:Fig5}(b) gives scattering rates which can be compared with rates obtained from the dc resistivity. In the $\omega\rightarrow 0$ limit the value of the $1/\tau$ is $3\times 10^{13}~{\rm s}^{-1}$ and $4\times 10^{13}~{\rm s}^{-1}$ for \bfca\ and \bfna, respectively. The classical expression for the conductivity $\sigma= ne^2\tau/m^*$ yields $1/\tau=3.6\times 10^{13}~{\rm s}^{-1}$ and $5\times 10^{13}~{\rm s}^{-1}$, correspondingly. This is in reasonable agreement with the values obtained from our extended Drude analysis. The calculated mean free path $\ell$, assuming a Fermi velocity $v_F= 2\times 10^6$~m/s of a typical metal \cite{Kittel} is of $50$~nm.

Up to now our approach was very conservative using only a classical concept (extended Drude analysis) which does not make assumptions that imply a microscopic interpretation. In the subsequent steps, we consider the electronic response due to two more-or-less independent electronic subsystems with some very distinct properties. Such an analysis is in part motivated by the multiband nature of iron-pnictides and by observation of two distinct superconducting gaps attributed to two distinct parts of Fermi-surface. Furthermore, such decomposition not only nicely fits the data but it is also the simplest one. Importantly, it leads to an unexpectedly simple description of the temperature dependent part of optical conductivity in terms of non or weakly interacting Fermi liquid.

\subsection{Decomposition of optical conductivity}
\label{sec:Decomposition}
From Fig.~\ref{fig:Fig6} it becomes obvious that the spectra of various 122 materials are very similar at elevated temperatures, in the normal state, no matter which ground state they eventually enter when cooled down. While the parent compounds \bfa, SrFe$_2$As$_2$, CaFe$_2$As$_2$ and \efa\ undergo a SDW transition below 200~K,\cite{Wu09,Hu08,Nakajima09} the electron doped materials \bfca\ and \bfna\ as well as the electron doped \bkfa\ become superconducting at $T_c=25$, 20 and 37~K, respectively.\cite{Wu09NP,Li08SC} The completely substituted BaNi$_2$As$_2$ is again non-superconducting.\cite{Chen09} There, the conductivity is significantly larger because of the increased carrier concentration; Ni donates twice as many electrons per ions compared to Cu.

Following the widely used procedure, we fit the complex optical conductivity $\hat{\sigma}(\omega) = \sigma_1(\omega)+{\rm i}\sigma_2(\omega)$ by a sum of Drude and Lorentz terms, keeping the number of contributions as small as possible:\cite{DresselGruner02}

\begin{eqnarray}
\hat{\sigma}(\omega)& =& \sum_{\rm Drude}\frac{N e^2 \tau}{m}
\frac{1}{1-{\rm i}\omega \tau}
\nonumber \\
& +&   \sum_{\rm Lorentz}\frac{N e^2}{m} \frac{\omega}{{\rm i}(\omega_0^2 - \omega^2)+\omega/\tau} \quad .
\label{eq:DrudeLorentz}
\end{eqnarray}

Here $N$ is the number of contributing carriers,
$e$ and $m$ are their electronic charge and mass. They define the plasma frequency: $\omega_p^2 = 4\pi N e^2/m$.
$\omega_0$ is the center frequency of the Lorentz oscillator,
$1/\tau$ is the relaxation rate describing the scattering or damping of the excitations.\cite{remark3}

\begin{figure}
 \centering
 \includegraphics[width=0.7\columnwidth]{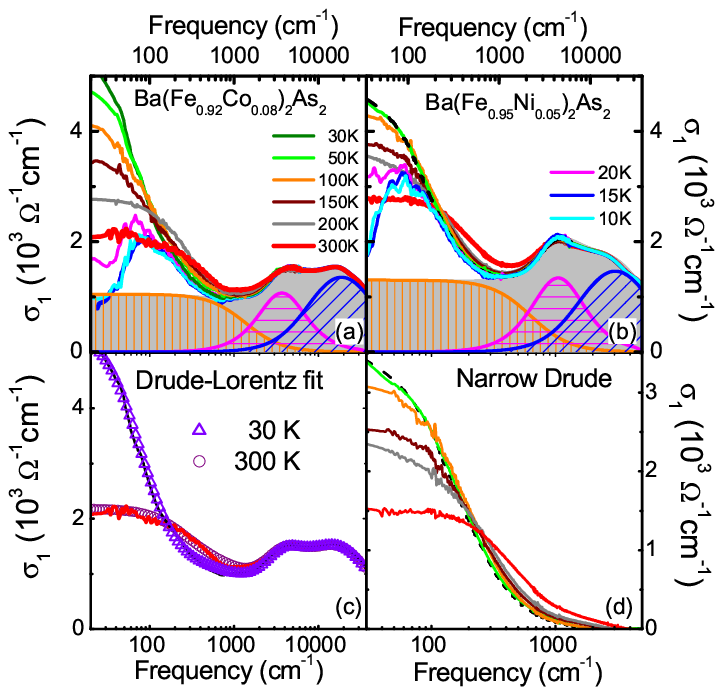}
 \caption{\label{fig:Fig8} (Color online) (a) and (b) Optical conductivity of \bfca\ and \bfna\ at temperatures between 10 K and 300~K in a wide spectral range. For both compounds the optical conductivity in the normal state can be separated in temperature dependent and independent parts. The temperature independent part consists from the incoherent broad term $\sigma_B$ (medium shaded orange), and two effective oscillators at higher frequencies (sparsely hatched magenta and green). Sum of the three contributions results in the gray shaded area. The unshaded area below the curves is temperature dependent and can be described by a single Drude term ($\sigma_N$) as discussed in the text. Quality of Drude-Lorentz fits is demonstrated for \bfca\ in the panel (c) for two temperatures $T=30$~K and 300~K. Temperature evolution of $\sigma_N$, in the normal state of \bfna\ is shown in the panel (d).}
\end{figure}

For all members of the 122 family there seems to be a constant background in the conductivity that does not change very much with temperature and composition. On the top of it there are additional contributions, which are specific for the different compounds and exhibit distinct temperature dependences. This is demonstrated in the Fig.~\ref{fig:Fig8} for two compounds, \bfca\ and \bfna\, which have been extensively studied here. The temperature independent part of the conductivity is indicated by the gray shaded area, and it consist at low frequencies from a single relatively broad Drude-like term $\sigma_B$ (medium hatched orange area) and two oscillators that mimic sufficiently well the high frequency contributions (magenta and blue hatched area). Significantly, the remaining temperature dependent part of conductivity Fig.~\ref{fig:Fig8}(d) in the normal state can be described with a single Drude term $\sigma_N$. The overall quality of the fit is shown in Fig.~\ref{fig:Fig8}(c) for the highest measured temperature and temperature just above the superconducting transition. It was shown that identical decomposition also holds for the normal state of \efa.\cite{Wu09}

However, already at elevated temperatures upon passing through the SDW transition in the case of \bfa, SrFe$_2$As$_2$, \efa, or BaNi$_2$As$_2$, for instance, a gap in the excitation spectrum opens since parts of the Fermi surface are removed. Consequently, spectral weight in $\sigma_N$ shifts from low-frequencies to above approximately 1000~\cm. Nevertheless, a small Drude-like peak (accounting for the metallic dc properties) remains and it becomes very narrow as the temperature is reduced.\cite{Wu09,Hu08,Nakajima09,Chen09} On the other hand, the metallic state of the superconductors close to optimal doping, such as \bfca, \bfna\ or \bkfa\ does not exhibit any trace of the SDW pseudogap. Upon cooling down to the superconducting transition, $\sigma_N (\omega)$ narrows and grows; spectral weight shifts to lower frequencies and the overall spectral weight [Eq.~\ref{eq:spectralweight}] is slightly reduced, as discussed in Sec.~\ref{sec:SpectralWeight}.

The suggested decomposition seems to be universal for the normal state of the whole 122 family directly, which follows from Fig. 6. The conductivity of nine different compounds (four parent, three electron doped, one hole doped and one in which the Fe is entirely substituted Ni) is decomposed in the above mentioned four terms, fitting satisfactorily the data. Furthermore, it is important to note that one of the spectra belongs to the 90 nm Ba\-(Fe$_{0.9}$\-Co$_{0.1})_2$\-As$_{2}$ thin film. Thus, we can rule out the flux contribution in the optical spectra. Consequently the existence of two electronic subsystems is most likely an intrinsic property of 122 systems. In addition, it is worth of mentioning that a recent study on the Co doped series of Ba(Fe$_{1-x}$Co$_x$)$_2$As$_2$, where $x$ is varied between 0 and 0.08, obey the suggested decomposition in the normal state regardless of doping.\cite{Uchida}

Notably, there seems to be no considerable transfer of spectral weight between both subsystems upon cooling within the error bars ($\approx$ 2$\%$ for each contribution) of our experiment and fits. We are inclined to ascribe the two low frequency Drude-like components to contributions from different bands crossing the Fermi surface. This immediately implies that they have a distinct influence on the electronic properties of these compounds. The coherent narrow component $\sigma_N$ contains only a quarter of the observed optic spectral weight compared to the incoherent background $\sigma_B$ but it turns out that it is of superior importance to their low energy behavior.

\subsection{Hidden Fermi liquid}
\label{sec:Hidden}

\begin{figure}
 \centering
\includegraphics[width=0.7\linewidth]{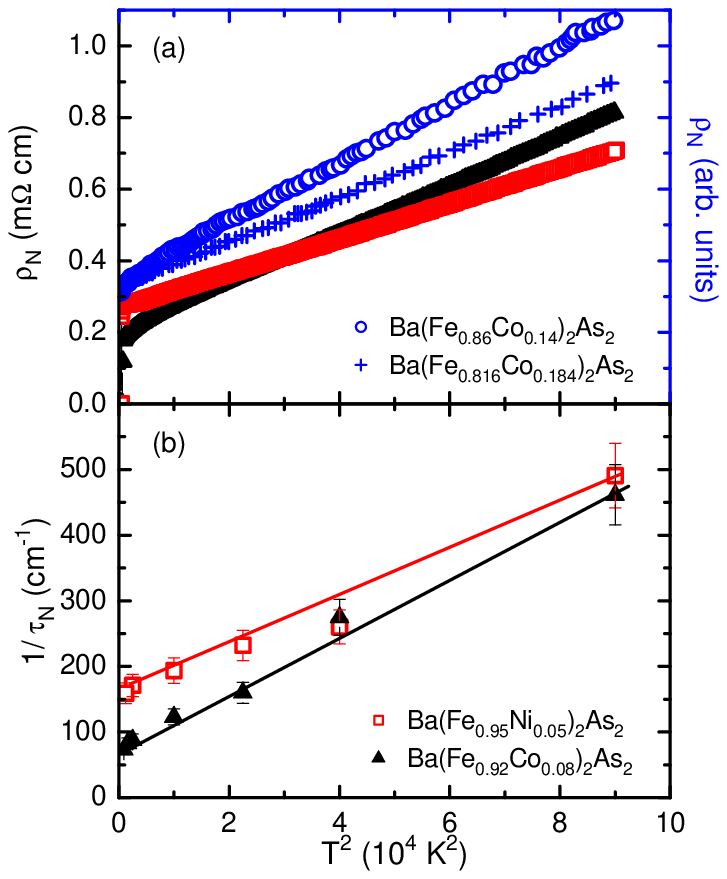}
 \caption{\label{fig:Fig7} (Color online)
(a) The narrow Drude-like contribution
obtained by decomposition by Eq.~(\ref{eq:decomposition}) plotted as a function of $T^2$. For a variety of ironpnictides
[\bfna\ (red squares), \bfca\ (black triangles), Ba(Fe$_{0.86}$Co$_{0.14}$)$_2$As$_2$ (blue circles)\cite{Chu09} and Ba(Fe$_{0.816}$Co$_{0.184}$)$_2$As$_2$ (blue pluses)\cite{Chu09} in the metallic phase a quadratic temperature dependence is observed all the way up to room temperature: $\rho_N\propto T^2$. Since in Ref.~\onlinecite{Chu09} no absolute values have been reported, the data displayed by blue dots correspond to the right scale in arbitrary units.
(b) Temperature dependence of the scatting rate $1/\tau_N(T)$ of Ba(Fe$_{1-x}M_{x}$)$_2$As$_2$ obtained from the fit of the low-frequency optical data by the Drude model Eg.~(\ref{eq:Drude}).}
\end{figure}

No question, the optical conductivity can be decomposed in several ways. The analysis presented here, however, is the simplest one;\cite{remark5} in addition a closer inspection of the temperature dependence of the narrow Drude component, as we will discussed now, makes obvious that some fundamental issues are uncovered, evidencing a deep physical consistency of our decomposition.

In order to reveal the temperature dependence of the dc resistivity of the subsystem described by $\sigma_N$, we follow the same logic as in the case of the optical conductivity. Consequently we decompose the conductivity at $\omega=0$ into the sum of $\sigma_N$ and $\sigma_B$, we restrain ourselves from assuming any temperature dependence for the later one:
\begin{equation}
\sigma_{\rm total}(T) = \sigma_N(T) + \sigma_B \quad .
\label{eq:decomposition}
\end{equation}
For each compound, we simply deduced $\sigma_B$ from the Drude-Lorentz fits displayed in Figs.~\ref{fig:Fig6} and \ref{fig:Fig8}. This procedure gives a surprising but very simple result, namely that
\begin{equation}
\rho_N(T) = \frac{1}{\sigma_N(T)}= \rho_0 + A_2\, T^2
\end{equation}
all the way up to room temperature, as shown in Fig.~\ref{fig:Fig7}(a). Importantly, no kink induced by a gap or pseudogap is observed in contrast to the parent compound, also suggesting that the SDW correlations are weak or absent in the samples close to the optimal doping.

We strongly feel that the observation of a simple power law $T^2$ in dc resistivity is not a pure coincidence. In fact it suggests strongly that the decomposition of the optical conductivity presented in Sec.~\ref{sec:Decomposition} is correct. An independent confirmation\cite{remark8} is obtained from the scatting rate $1/\tau_N(T)$ of Ba(Fe$_{1-x}M_x)_2$As$_2$: fitting the optical conductivity $\sigma_N(\omega)$ by the Drude model
\begin{equation}
\hat{\sigma}_N(\omega) = \frac{N e^2 \tau_N}{m}
\frac{1}{1-i\omega \tau_N}
\label{eq:Drude}
\end{equation}
yields $1/\tau_N(T)\propto T^2$, as presented in Fig.~\ref{fig:Fig7}(b). Note, both approaches do not just get the same temperature dependence, also the slope for \bfca\ is always larger compared to \bfna. From our fit we obtain $6.64\times 10^{-9}~\Omega{\rm cm K}^{-2}$ for \bfca\ and $5 \times 10^{-9}~\Omega{\rm cm K}^{-2}$ for \bfna.

The $T^2$ dependence of scattering rate and the dc resistivity is generally associated with a Fermi liquid. Examples of similar temperature dependence of dc resistivity up to elevated temperatures can be found in: Ruthenate superconductors,\cite{Maeno97} organic layered superconductors,\cite{Bulaevskii88} transition metal dichalcogenides,\cite{Julien91} High temperature superconductors.\cite{Nakamae03,Taillefer09,Manako92,Alloul89} In Fermi liquids the weak Umklapp electron-electron scattering processes are dominant in the scattering rate which corresponds to
\begin{equation}
\hbar/\tau_e(T) =A(k_BT)^2/E_F \quad ,
\label{eq:FL}
\end{equation}
where $A$ is a dimensionless constant. Using the same numbers for the Fermi energy as in Ref.~\onlinecite{Rullier09}, we found that $A\approx 1-2$. This is consistent with the weak coupling limit behind the above expression, which assumes that the effective Umklapp is small with respect to the Fermi energy. It is worth of noting that a possibly relevant complication in estimating $E_F$ occurs for strongly interacting electron systems. The term ``normal Fermi system'' should then refer only to quasi particles which obey the Fermi-Dirac statistics. This produces an ambiguity in the definition of the appropriate $E_F$.

Notably, the proximity of the SDW and the SC phase, above which the Fermi liquid state is found, evokes the phase diagram of the organic superconductors. Those materials are understood in terms of Umklapp process, normal scattering and nesting.\cite{Emery} If the Umklapp process is weak with respect to $E_F$, Fermi liquid behavior follows from Eq.~(\ref{eq:FL}) at low temperatures. Nevertheless, as well known, though the normal electron-electron scattering makes no contribution to the resistance, when attractive it can produce superconductivity as in the case of organic conductors. On the other side, if Umklapp is enhanced due to nesting, the system scales towards the strong coupling regime in which SDW is a possible ground state. Such reasoning opens an interesting avenue for understanding of iron-pnictides.

Our finding of a hidden $T^2$ behavior is in qualitative agreement with the conclusion drawn from comprehensive Hall-effect studies on the series \BaFeCoxAs\ by Rullier-Albenque {\it et al.} \cite{Rullier09} It was pointed out there for the first time that the iron-pnictides are archetypical Fermi liquids. However, it should be mentioned that our way to a hidden $T^2$ differs from that followed in the analysis of the Hall coefficient where a change in electron concentration of approximately $20-30\%$ was assumed with temperature. In contrast, our decomposition of optical conductivity reveals that the spectral weight of $\sigma_N$ remains unchanged to the precision of the experiment, which is less than a few percent. Furthermore it seems that the coefficient A$\approx$4 estimated\cite{Rullier09} from Hall-effect seems somewhat large to be consistent with the weak coupling limit.

It is also worth mentioning that doping does not change the slope of the resistivity curves, but only the intercept $\rho_0$.\cite{Chu09,Rullier09} Thus it can be asserted that the Fermi-liquid behavior extends through the complete doping series, i.e.\ the whole paramagnetic phase of these compounds. Furthermore other iron-pnictides behave in a very similar way, as demonstrated on \bfna\ in Fig.~\ref{fig:Fig7}(b), confirming our general view. Notably, a $T^2$ behavior of the resistivity is also found in 1111 compounds\cite{sefatprb,Suzuki09} while a recent optical conductivity study revealed the Fermi-liquid behavior in FeTe$_{0.55}$Se$_{0.45}$ compound\cite{Homes10}. However, while the $A_2$ values extracted in 1111 compound are comparable with moderate heavy-fermion compounds, such as CePd$_3$ and UIn$_3$,\cite{KadowakiWoods86,Jacko09} we obtain values one order of magnitude smaller in the 122 family. This places the 122 iron-pnictides right between the cuprates and conventional metals, as far as the correlation strength is concerned. Our finding is in very good accord with conclusions drawn from comparison of the spectral weight with band structure calculations.\cite{Qazilbash09,Mazin09}

\begin{figure}
 \centering
 \includegraphics[width=0.7\columnwidth]{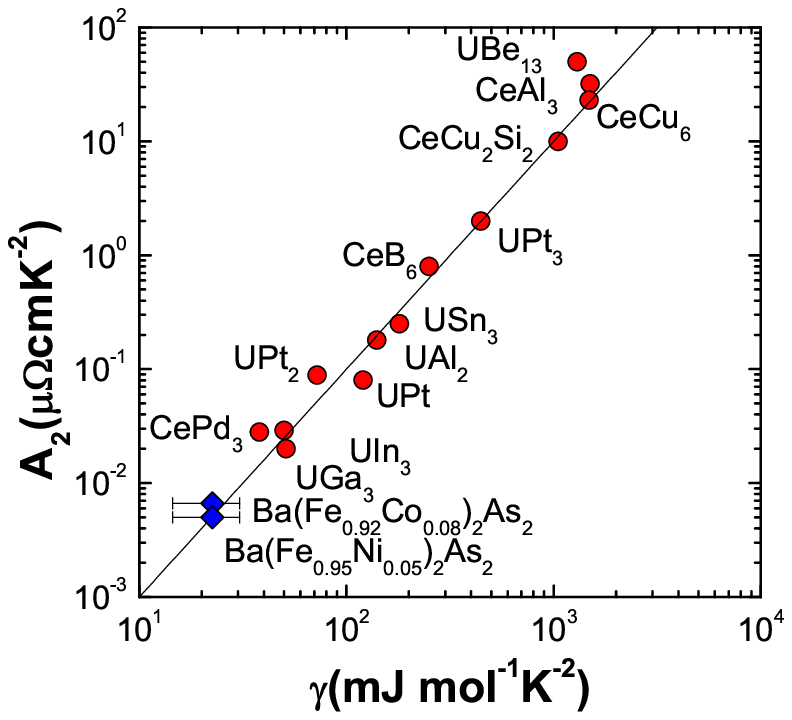}
 \caption{\label{fig:KadowakiWoods} (Color online) Original points in the Kadowaki-Woods plot \cite{ KadowakiWoods86} for heavy and "semi-heavy" fermion systems are indicated by the red circles. The points for \bfna\ and \bfca\ are indicated by blue squares.}
\end{figure}

Since 122 compounds exhibit a Fermi liquid behavior it is of interest to check whether the empirical Kadowaki-Woods relation is obeyed. This relation links the resistivity and the specific heat giving universally a constant ratio $A_2/\gamma^2= 1.0 \cdot 10^{-5}\mu\Omega$cm(mole K/mJ)$^2$. $\gamma$ is the Sommerfeld coefficient of the linear term in the temperature dependence of the specific heat. Typically in iron-pnictides it varies between 7 and 35 mJ/moleK$^2$.\cite{Welp09,Rotter08,Ni08} As shown in Fig.~\ref{fig:KadowakiWoods} the agreement between the universal relation and the data is remarkable which only reinforces further our conclusions.

\section{Conclusions}

We have measured the electrodynamic properties of \bfca\ and \bfna\ single crystals in a wide frequency range as a function of temperature and presented a detailed analysis of their behavior in the metallic state above $T_c=25$ and 20~K, respectively.
In a one-component picture, the extended Drude analysis yields the frequency dependence of the effective mass and scattering rate. $m^*/m_b$ is enhanced at low temperatures and frequencies by approximately a factor of 5. The low-frequency scattering rate $1/\tau(\omega)$ decreases on cooling but turns into the opposite behavior above 400~\cm. The spectral weight shifts to lower energies with decreasing temperature; a small but significant fraction is not recovered within the infrared range of frequencies.

Comparing the optical spectra a numerous 122 iron-pnictides, it becomes obvious that the frequency-dependent conductivity can be universally decomposed into a broad temperature-independent background $\sigma_B$ and a narrow Drude-like component $\sigma_N(T) $ which solely determines the transport properties.
The conducting subsystem $\sigma_N(T)$ reveals a $T^2$ behavior in the dc resistivity up to elevated temperatures, indicating a hidden Fermi liquid behavior of the these iron-pnictide superconductors. We conclude that the superconducting state evolves out of a Fermi liquid state where correlations are important to a moderate extend; we find no indications of a quantum critical behavior.


\begin{acknowledgments}
We thank D.N. Basov, N. Drichko, B. Gorshunov and N.L. Wang for helpful discussions. The
contributions of J. Braun, A. Faridian, P. Kallina, X. Lin and E.S. Zhukova to the
experiments are appreciated. D.W and N.B. acknowledge their
fellowship by the Alexander von Humboldt Foundation.
The work at Zhejiang
University was funded by NSF of China.
\end{acknowledgments}


\begin{thebibliography}{99}
\bibitem{remark1}
There are indications that some details of the optical properties might actually depend in the particular system under investigation.
\bibitem{remark4}
The room temperature resisitivity is very close to $\rho_{ab}=400~\mu\Omega$cm of the undoped parent compound \efa, measured previously.\cite{Wu09} In that compound, however, the initial slope $|{\rm d}\rho/{\rm d}T|$ is smaller.
\bibitem{remark2}This is supported by $c$-axis reflectivity measurements on \efa\ (Ref.~\onlinecite{Wu09}) where the interband transition is also seen.
\bibitem{remark3}
For our analysis, the reflectivity and conductivity are fitted
simultaneously.
\bibitem{remark5}Other decompositions of the optical conductivity than suggested
here might be tempting, too. For example, the next simplest one is
to fit the low-frequency conductivity by a Drude and a Lorentzian
terms. However, there is not really any physical meaning of the
Lorentzian, and furthermore its position, width and amplitude
strongly change with temperature which is highly anomalous.
\bibitem{remark8}
Although we have utilized the dc resistivity for the extrapolation
of the low-frequency reflectivity, the determination of the Drude
roll-off in the optical conductiivty is to a large extend
independent from this procedure.
\end{thebibliography}
\end{document}